\documentstyle[12pt,epsfig]{article}

\advance\textheight by \topskip
\begin{document}
\tolerance=100000
\thispagestyle{empty}
\setcounter{page}{0}

\newcommand{\be}{\begin{equation}}
\newcommand{\ee}{\end{equation}}
\newcommand{\br}{\begin{eqnarray}}
\newcommand{\er}{\end{eqnarray}}
\newcommand{\ba}{\begin{array}}
\newcommand{\ea}{\end{array}}
\newcommand{\bi}{\begin{itemize}}
\newcommand{\ei}{\end{itemize}}
\newcommand{\bn}{\begin{enumerate}}
\newcommand{\en}{\end{enumerate}}
\newcommand{\bc}{\begin{center}}
\newcommand{\ec}{\end{center}}
\newcommand{\ul}{\underline}
\newcommand{\ol}{\overline}
\newcommand{\eennH}{$e^+e^-\rightarrow  \bar\nu_e\nu_e H$}
\newcommand{\eennbb}{$e^+e^-\rightarrow  \bar\nu_e\nu_e b\bar b$}
\newcommand{\eennbbph}{$e^+e^-\rightarrow  \bar\nu_e\nu_e b\bar b\gamma$}
\newcommand{\Hbb}{$H \rightarrow  b\bar b$}
\newcommand{\Hbbph}{$H\rightarrow  b\bar b\gamma$}
\newcommand{\eennjjjj}{$e^+e^-\rightarrow \bar\nu_e\nu_e \mathrm{jjjj}$}
\newcommand{\eennjjjjph}{$e^+e^-\rightarrow\bar\nu_e\nu_e\mathrm{jjjj}\gamma$}
\newcommand{\Hjjjj}{$H\rightarrow \bar\nu_e\nu_e \mathrm{jjjj}$}
\newcommand{\Hjjjjph}{$H\rightarrow \bar\nu_e\nu_e \mathrm{jjjj}\gamma$}
\newcommand{\eezh}{$e^+e^-\rightarrow ZH$}
\newcommand{\uub}{$ u\bar u$}
\newcommand{\ddb}{$ d\bar d$}
\newcommand{\ssb}{$ s\bar s$}
\newcommand{\ccb}{$ c\bar c$}
\newcommand{\bbb}{$ b\bar b$}
\newcommand{\ttb}{$ t\bar t$}
\newcommand{\eeb}{$ e^+ e^-$}
\newcommand{\mumub}{$ \mu^+\mu^-$}
\newcommand{\tautaub}{$ \tau^+\tau^-$}
\newcommand{\veveb}{$ \bar\nu_e\nu_e$}
\newcommand{\vmvmb}{$ \bar\nu_\mu\nu_\mu $}
\newcommand{\vtvtb}{$ \bar\nu_\tau\nu_\tau $}
\newcommand{\lra}{\leftrightarrow}
\newcommand{\ar}{\rightarrow}
\newcommand{\sm}{${\cal {SM}}$}
\newcommand{\MH}{M_{H}}
\newcommand{\MW}{M_{W}}
\newcommand{\MZ}{M_{Z}}
\newcommand{\Dir}{\kern -6.4pt\Big{/}}
\newcommand{\Dirin}{\kern -10.4pt\Big{/}\kern 4.4pt}
\newcommand{\DDir}{\kern -7.6pt\Big{/}}
\newcommand{\DGir}{\kern -6.0pt\Big{/}}
\def\Ord{\buildrel{\scriptscriptstyle <}\over{\scriptscriptstyle\sim}}
\def\OOrd{\buildrel{\scriptscriptstyle >}\over{\scriptscriptstyle\sim}}
\def\pl #1 #2 #3 {{\it Phys.~Lett.} {\bf#1} (#2) #3}
\def\np #1 #2 #3 {{\it Nucl.~Phys.} {\bf#1} (#2) #3}
\def\zp #1 #2 #3 {{\it Z.~Phys.} {\bf#1} (#2) #3}
\def\pr #1 #2 #3 {{\it Phys.~Rev.} {\bf#1} (#2) #3}
\def\prep #1 #2 #3 {{\it Phys.~Rep.} {\bf#1} (#2) #3}
\def\jp #1 #2 #3 {{\it J.~Phys.} {\bf#1} (#2) #3}
\def\prl #1 #2 #3 {{\it Phys.~Rev.~Lett.} {\bf#1} (#2) #3}
\def\mpl #1 #2 #3 {{\it Mod.~Phys.~Lett.} {\bf#1} (#2) #3}
\def\rmp #1 #2 #3 {{\it Rev. Mod. Phys.} {\bf#1} (#2) #3}
\def\xx #1 #2 #3 {{\bf#1}, (#2) #3}
\def\preprint{{\it preprint}}

\begin{flushright}
{\large DFTT 26/96}\\ 
{\large Cavendish--HEP--96/07}\\ 
{\rm May 1996\hspace*{.5 truecm}}\\ 
\end{flushright}

\vspace*{\fill}

\begin{center}
{\Large \bf Higgs signals and hard photons via ${WW}$-fusion\\
in the Standard Model\\  
at the Next Linear Collider}\\[0.95cm]
{\large Stefano Moretti\footnote{E-mails: Moretti@to.infn.it; 
Moretti@hep.phy.cam.ac.uk.}}\\[0.25cm]
{\it Dipartimento di Fisica Teorica, Universit\`a di Torino,}\\
{\it and I.N.F.N., Sezione di Torino,}\\
{\it Via Pietro Giuria 1, 10125 Torino, Italy.}\\[0.25cm]
{\it Cavendish Laboratory, 
University of Cambridge,}\\ 
{\it Madingley Road,
Cambridge, CB3 0HE, United Kingdom.}\\[0.2cm]
\end{center}
\centerline{PACS numbers: 14.80.Bn, 14.70.Bh, 14.70.Fm, 29.17.+w.}
\vspace*{\fill}

\begin{abstract}
{\small
\noindent
Within the framework of the Standard Model, integrated and differential
distributions are given for Higgs production via the $WW$-fusion mechanism and
decay via the channels $H\ar b\bar b$ and $H\ar WW\ar \mathrm{jjjj}$,
with and without photon radiation, at Next Linear Collider energies.
Calculations are carried out at tree-level and rates 
of the leading processes 
$e^+e^-\rightarrow\bar\nu_e\nu_e H \ar \bar\nu_e\nu_e b\bar b $ and
$e^+e^-\rightarrow\bar\nu_e\nu_e H \ar \bar\nu_e\nu_e WW 
\ar \bar\nu_e\nu_e \mathrm{jjjj}$ are compared to those
of the next-to-leading reactions
$e^+e^-\rightarrow\bar\nu_e\nu_e H (\gamma)\ar \bar\nu_e\nu_e b\bar b \gamma$
and $e^+e^-\rightarrow\bar\nu_e\nu_e H (\gamma)\ar \bar\nu_e\nu_e WW (\gamma)
\ar \bar\nu_e\nu_e \mathrm{jjjj}\gamma$, 
in the case of hard and detectable photons.
Finally, a brief discussion concerning the case of $H\ar ZZ\ar\mathrm{jjjj}
(\gamma)$ decays
is also given.}
\end{abstract}

\vspace*{\fill}
\newpage
\section*{1. Introduction}

The motivations for building an $e^+e^-$ linear collider operating in the
energy range $\sqrt s=300-1000$ GeV (NLC, Next Linear Collider)
are indeed quite convincing 
\cite{epem1,epem2,epem3,epem4,epem5}. 
For example, one of the major goals of such a machine
will be to perform `high precision' Higgs physics. That is, to measure
the parameters (mass $\MH$, width $\Gamma_H$, spin ${\cal S}_H$,
$\{{\cal J^{PC}}\}_H$ quantum numbers, etc.) and the relevant
phenomenological rates (cross sections, branching ratios, etc.)
of the Higgs boson $H$, to an accuracy that can not be achieved 
at hadron machines.
The importance of this project is clear if one considers that the
Higgs mechanism of spontaneous symmetry breaking is a sort of `pedestal'
that holds up the entire Standard Model (\sm).

If it exists, the $H$ particle will be probably discovered at the Large
Hadron Collider (LHC)
\cite{LHC}, which is expected to start running
early next century, well in advance of the future NLC. 
The CERN $pp$ collider will however have great difficulties in measuring
the fundamental quantities related to this particle, because of
the huge background proceeding via strong interactions which will be present
at the hadron machine.
This is particularly true for a Higgs boson in the intermediate
mass range (IMR), $\MH\Ord2\MW$, for which the detectability 
of the only two viable decay channels, 
$H\ar\gamma\gamma$ and $H\ar b\bar b$, strongly depends on the detector
performances. For a Higgs in the heavy mass range (HMR), 
$\MH\OOrd2\MW$, things should be 
easier, as the four-lepton decay mode $H\ar ZZ\ar 4\ell$ is relatively
clean and straightforward to detect.

At the NLC, the main \sm\ Higgs production mechanisms are 
via the bremsstrahlung process \eezh\ \cite{bremSM} and via the
$WW$- and $ZZ$-fusion reactions $e^+e^-$ $\rightarrow 
\bar\nu_e\nu_eW
W(e^+e^-ZZ)$$ \rightarrow\bar\nu_e\nu_e
(e^+e^-)H$ \cite{fusionSM}. 
For a first stage NLC (with $\sqrt s\Ord$ 500 GeV) the rates of the 
bremsstrahlung
mechanism are larger than those of the fusion channels, if $\MH\Ord2\MW$.
At larger centre-of-mass (CM)
 energies ($\sqrt s\OOrd 500$ GeV) $WW$-fusion 
starts dominating over the whole of the $\MH$ spectrum. Furthermore, the rates
of $ZZ$-fusion are generally one order of magnitude smaller than those
of the $WW$-channel. Concerning possible Higgs signatures, 
all the principal decay modes of this particle 
can be detected and studied at the NLC, for all values
of $\MH$, provided that enough statistics can be accumulated \cite{DHKMZ}.
In particular, the main sources of Higgs events will be the channels 
$H\ar b\bar b$ (IMR) and $H\ar WW\ar 4\mathrm{jet}$ (HMR) 
\cite{GHS}.

There are a few important aspects, in the way a NLC running 
around the TeV energy scale will operate,
that should be carefully considered in order
to exploit in full its potential and that are 
absent in lower energy \eeb\ machines.
These are related to the influence on the cross sections of 
electromagnetic (EM) interactions which can take place 
before the actual beam collision, such as {\sl bremsstrahlung}
and {\sl beamsstrahlung} effects \cite{ISR}\footnote{Effects due to
the energy spread of the beams before annihilation  (intrinsic to any 
collider) should also be considered.}.
Whereas effects due to synchrotron radiation emitted by one of the colliding
bunches in the field of the other one (i.e., beamsstrahlung) 
necessarily need, in order 
to be quantified, the knowledge of the technical details
of the collider design and can be realistically estimated only through
Monte Carlo simulations,
those due to the emission and exchange of photons from 
and between the actual pair of electron and positron which collide
(i.e., bremsstrahlung or Initial State Radiation, ISR),
can be treated in a fairly general way. In many cases, 
one can compute the exact EM corrections to the $e^+e^-$ 
annihilation subprocess and express these via the
so-called `electron structure functions'.
Such corrections embody both real and virtual photon radiation
and they are known to date up to the order ${\cal O}(\alpha_{\mathrm{em}}^2)$
\cite{structure}.
It has also been shown that, for `narrow beam' designs, beamsstrahlung
affects the cross section much less than the ISR \cite{ISR}, such that in 
phenomenological analyses one can consistently confine oneself to 
dealing with bremsstrahlung radiation only.  
In general, the principle effect of the ISR is to lower the effective 
CM energy available in the main process, thus ultimately
reducing(enhancing) total cross sections which  
increase(decrease) at larger CM energies.
Furthermore, ISR  also leads to a smearing of the 
differential distributions \cite{BCDKPZ}. However, the 
`electron structure functions' approach is not always applicable, such
as in the context of $WW$-fusion processes in $e^+e^-$ annihilations.

It is the purpose of this paper to study the properties
and the effect on the integrated rates as well as on the
differential distributions of interest to Higgs searches
of {\sl hard photon} emission,
which can take place in the \sm\ Higgs production and decay processes
\be\label{bbph} 
e^+e^-\rightarrow\bar\nu_e\nu_e H (\gamma)\ar \bar\nu_e\nu_e b\bar b \gamma,
\ee
\be\label{WWph} 
e^+e^-\rightarrow\bar\nu_e\nu_e H (\gamma)\ar \bar\nu_e\nu_e WW (\gamma)
\ar \bar\nu_e\nu_e \mathrm{jjjj}\gamma,
\ee
(that is, via the $WW$-fusion mechanism),
at NLC energies and for several values of the Higgs mass, and where
the photon can be emitted from the initial, virtual and 
final states.  
In our opinion, the importance of this study is motivated by the fact that,
contrary to lower energy $e^+e^-$ colliders (such as LEP1, SLC and in part
also LEP2), which `sit' on gauge boson resonances and so 
the width of the unstable particles imposes a natural cut-off  on events
with hard photons produced by the initial state,
at the  NLC such a suppression does not act any longer.
In addition, as the beam energy is much larger, the probability
that the incoming electrons and positrons can radiate hard photons
increases greatly. From this, 
it follows that in practise a sample of pure \eennbb\ and \eennjjjj\ 
events, without $\gamma$-radiation,
does not exist and one inevitably has to deal with
EM emission. Moreover, in the experimental samples of data, events in which
the photon is emitted during the production process 
$e^+e^-\ar \bar\nu_e\nu_e H\gamma$
are not distinguishable from those
in which the radiation comes from the Higgs decay stages $H\ar b\bar b\gamma$
and $H\ar WW(\gamma)\ar \mathrm{jjjj}\gamma$,
such that in phenomenological simulations one is forced to consider 
both the cases at the same time. In particular, we stress that it is 
principally
the second kind of radiation which could spoil the shape of the 
Higgs resonances. 
 
The plan of the paper is as follows. In Sec.~2 we devote some
space to illustrate the computational techniques that we have used as well
as the numerical values adopted.
Sec.~3 presents our results, and in Sec.~4 we give a summary and
report the main conclusions to be drawn from this study.
                                                          
\section*{2. Calculation} 

The Feynman diagrams corresponding to processes (\ref{bbph})--(\ref{WWph})
are given in Figs.~1 and 2, respectively. To compute them we have
adopted numerical techniques which use the helicity amplitude formulae
of Ref.~\cite{HZ} and the subroutines contained in the package
HELAS \cite{HELAS}. For process (\ref{bbph}) we have also resorted to
the program MadGraph \cite{tim} for the generation of the {\tt FORTRAN} code
of the Matrix Element (ME). The routine 
{\tt VEGAS} \cite{VEGAS} has been used as multi-dimensional integrator. 
To achieve high accuracy in the
evaluation of the cross sections in presence of Higgs peaks
in different regions of the phase space,
we have used the method of splitting the Feynman amplitudes squared
into non-gauge invariant terms, each of which with a different
peak structure. These have been then integrated separately 
by using an appropriate choice of the phase space variables,
according to the resonant
behaviour of the graphs involved. A sum over the various
terms of the MEs gives in the end gauge-invariant results. Such a procedure
has been already described elsewhere (for example, in Ref.~\cite{cave9517}),
such that we do not enter here into details. 

The computation of the graphs in Figs.~1--2 is rather straightforward,
though the numerical integrations over the phase space can  in some 
instances be rather complicated, as they can involve up to 16 
dimensions\footnote{Corresponding to a 
final state of 7 particles, in case of process (\ref{WWph}).}.
Thus, in order to perform our calculations in a reasonable amount of
CPU time we have always constrained the $W$-bosons to be on-shell in process
(\ref{WWph})\footnote{Both in the case of process ({\ref{bbph}}) 
and in the lowest order reaction
\be\label{bb} 
e^+e^-\rightarrow\bar\nu_e\nu_e H \ar \bar\nu_e\nu_e b\bar b,
\ee
no simplification in any respect has been adopted.}. 
In this way, we are able to reduce by two the number of the variables
we integrated over. As we will be eventually interested in studying the 
invariant mass distributions of the decay products of the Higgs boson, we
expect such a procedure to have a little impact on our final results. 
However, a direct consequence of this simplification is that we are not able
to study Higgs decays into four jets below and near the $2\MW$ threshold.
Nonetheless, since the phenomenology of the off-shell Higgs decay
$H\ar W^*W^*$ is well known, predictions in the range
$\MH\Ord2\MW$ can be easily extrapolated from our rates. 
For consistency, we have constrained the $W$-bosons to be on-shell also in 
the case of the leading process 
\be\label{WW} 
e^+e^-\rightarrow\bar\nu_e\nu_e H \ar \bar\nu_e\nu_e WW 
\ar \bar\nu_e\nu_e \mathrm{jjjj}.
\ee
Furthermore, the interferences between the various peaks in
processes  (\ref{bbph})--(\ref{WWph}) are in general very small compared 
to the squared terms of the resonances and do not bring
any distinctive structure into the differential distributions. 
In fact, such contributions always mix up
graphs with different resonant structures\footnote{We have integrated them
by using a flat phase space, which does not map any of the possible
peaks of the interfering graphs.}, such that the
phase space regions in which one or more amplitudes are large are 
different for different graphs.
Therefore, as a further simplification, such interferences
have been systematically 
neglected in the presentation of our results.

The following numerical values of the parameters have been adopted:
$M_{Z}=91.175$ GeV, $\Gamma_{Z}=2.5$ GeV,
$M_{W}=80.23$ GeV, and for the
Weinberg angle we have used its leptonic effective value 
$\sin^2_{\mathrm{eff}} (\theta_W)=0.2320$.
For the fermions: $m_e=m_{\nu_e}$=0, $m_b=4.25$
GeV, whereas all light quarks $u,d,s$ and $c$ have been considered
massless. Jets have been identified with the partons from 
which they originate. The {EM} coupling
constant $\alpha_{\mathrm{em}}$ has been set equal to 1/128. For
the Higgs width $\Gamma_{H}$ we have adopted the tree-level
expression corrected for the running of the quark masses in the vertices
$Hq\bar q$ (these have been
 evaluated at the scale $\mu=M_{H}$
\cite{running}). Therefore, in order to be consistent, we have used
a running $b$-mass in the $H\ar b\bar b$ vertex of the production
processes (\ref{bbph}) and (\ref{bb}). 
As representative values of the CM energy of 
the NLC we have adopted 300, 500 and 1000 GeV, 
and the Higgs mass has been chosen in the range 
$60~\mathrm{GeV}\Ord\MH\Ord450~\mathrm{GeV}$.  

\section*{3. Results}

Photon radiation can be emitted in processes (\ref{bbph})--(\ref{WWph}) 
both during the production mechanism \eennH$\gamma$\ and during the decays 
\Hbb$\gamma$\ and \Hjjjj$\gamma$. 
The emission is gauge-invariant in both these two stages
and we will refer to it as `production radiation' and `decay radiation',
respectively. The ISR as previously defined would correspond 
here to the photons emitted by the incoming electron and positron lines.
The diagrams describing such bremsstrahlung (i.e., graphs
1--2 in both Fig.~1--2) do not satisfy 
gauge-invariance if taken on their own. To recover the latter 
it is needed to add the graphs in which the photon is emitted by the
$W$-lines (i.e., graphs 5--6 in both Fig.~1--2). Therefore,  
a separation of the ISR diagrams from the rest does not
make sense here. The issue of defining the ISR in a gauge-invariant way 
in presence of charged current (CC) interactions is indeed  a well known 
problem, for example in case of $e^+e^-\ar W^+W^-\ar 4~\mathrm{fermions}$ 
production at LEP2.
In the sense that, in presence of electric charge flow between the initial and 
the subsequent stages of $e^+e^-$-initiated processes, the definition of the
ISR is not unique. Such a difficulty has been probably overcome in the case
of neutrino exchange, by means of the so-called {\sl current splitting 
technique} (for details, see Ref.~\cite{CST}). In that case, complete ISR
separates into a universal, factorising, process independent-contribution
(that is, the electron structure functions 
of Ref.~\cite{structure}) and a non-universal, non-factorising, 
process-dependent part (that is, the complex analytical expressions as
given in Ref.~\cite{ISRcomplete}). However, this approach is not realistically 
possible in the present context. In fact,
the formulae of Refs.~\cite{structure,ISRcomplete} are valid
only in case of {\sl annihilation}- and {\sl conversion}-type
Feynman diagrams, that is, when the incoming $e^+$- and $e^-$-lines 
are connected to each other via a
$s$- or $t,u$-channel, respectively.
In contrast, we are concerned here with electron/positron lines that are
disconnected from each other 
and that end up as the (anti)neutrino ones of the final states, by 
means of multiple space-like CC-interactions.

The most correct approach would certainly be to compute the
complete ${\cal O}(\alpha_{\mathrm{em}})$ (and beyond) corrections, 
including the loop
diagrams. That is however beyond the intentions of this study.
What we will do here is to show results for the leading
order (LO) processes (\ref{bb})--(\ref{WW}) and the next-to-leading order (NLO)
ones (\ref{bbph})--(\ref{WWph}) separately, the latter with real photons 
having $p_T^\gamma>1$ GeV (hard and detectable EM radiation), in terms of
both integrated and differential rates.
We also impose an additional cut on the photon, by asking that the latter be
isolated from the jets arising in the final states of processes
(\ref{bbph})--(\ref{WWph}), from the $b$- and the 
light quarks: for example, by imposing
$\cos\theta_{b\gamma,~\mathrm{j}\gamma}<0.95$ (corresponding to
a cone with an angular size of $\approx18$ degrees). 
This is done because it would generally be
impossible to tag photons too close to the original parton,
as the latter gives rise to a jet with a finite angular size, such 
that if the photon
fails within the corresponding cone it will not be distinguished from the
other parts of the jets. In this case, 
its energy is counted as part of
that of the hadronic system associated with the parton
and the $M_{b\bar b}$ and $M_{\mathrm{jjjj}}$ invariant masses
are not experimentally measurable in 
events of the type (\ref{bbph})--(\ref{WWph}). For configurations
in which the photon is highly collinear (i.e., 
$\cos\theta_{b\gamma,~\mathrm{j}\gamma}\ge0.95$),  such
events would rather be recognised as  
leading $2\ar4$ and $2\ar6$ topologies.

In order to be consistent with the sketched approach, 
we will refrain from summing up lowest and next-to-lowest rates through the 
orders
${\cal O}(\alpha_{\mathrm{em}}^5)$ and ${\cal O}(\alpha_{\mathrm{em}}^7)$,
as such a summation would need to include also
the contributions due to virtual photons.
In practise then, the aim of our study is, on the one hand, to
estimate the size of the rates produced by events with hard photons and, on
the other hand, to compare their kinematic properties to those of the
non-radiative processes. This is done 
in order to establish whether the shape of the
differential spectra of interest to Higgs searches can be significantly
modified at higher order by the presence of hard photons, such that when 
proceeding to experimental analyses one might expect a
smearing of the relevant resonant distributions. In particular, 
we remind the reader that the contributions that we have not computed
(the loop diagrams) or removed by the cuts (the infrared photon regions) 
would show the same kinematics as the lowest order processes, such that they 
would modify the overall normalisation of the differential distributions but 
not the shape. 
In particular, concerning the `decay radiation', it is well known that the 
Kinoshita-Lee-Nauenberg theorem \cite{KLN} prescribes that 
all the logarithmic contributions of the form 
$(\alpha_{\mathrm{em}}/\pi)^n{\ln}^n(s/m^2_{b,~\mathrm{j}})$, 
which appear to the order ${\cal O}(\alpha_{\mathrm{em}})^n$
according to the Sudakov theorem \cite{Sudakov} and which are
due to `collinear' emission of photons from the final state, 
must not appear in the  expression of the inclusive cross section. They
are in fact canceled by the {\sl negative} contributions due to
virtual final state photons. Therefore, the
hard photon effects that we will discuss in the following
could well be larger in the
complete ${\cal O}(\alpha_{\mathrm{em}})$ result.
In the very end, in order to compare theory and experiment, the 
latter are certainly needed. However, our preliminary results
will enable us to assess whether, at higher order, complications
should be expected in individuating the position of the peaks and/or in
establishing their line-shape,
or whether the effect of ${\cal O}(\alpha_{\mathrm{em}})$ 
corrections is mainly matter of overall normalisation of the 
leading order distributions.

In our analysis, we will present spectra in
$M_{b\bar b}$, $M_{b\bar b\gamma}$, 
$M_{\mathrm{jjjj}}$, $M_{\mathrm{jjjj}\gamma}$ (that is,
in the invariant masses of the decay products $b\bar b(\gamma)$ and 
$\mathrm{jjjj}(\gamma)$ of the Higgs scalar), together with those
in $E_\gamma$ (the energy of the photon) and
in $p_T^{\mathrm{miss}}$ (the missing transverse momentum of the 
undetected neutrino pair).
The importance of the distributions in invariant mass is clear if one considers
that the presence of `decay radiation' could spoil in the end
the form of the Breit-Wigner
peaks in  $M_{b\bar b}$ and $M_{\mathrm{jjjj}}$
that one obtains from the LO contributions
(\ref{bb})--(\ref{WW}), via diagrams 3--4 in Fig.~1
and 3--4 \& 7--10 in Fig.~2. On the other hand, these
diagrams contribute to produce Higgs resonances in the 
$M_{b\bar b\gamma}$ and  $M_{\mathrm{jjjj}\gamma}$ spectra.
In contrast, the EM emission that takes place via the `production radiation'
should not significantly spoil the form of the peaks\footnote{Its effect
being mainly confined to lower the invariant mass of the $WW$-annihilation
into the Higgs boson.} and 
should instead contribute to enhance
the number of resonant events. 
The distributions in $E_\gamma$ and $p_T^{\mathrm{miss}}$  could in
principle be useful in order to separate
in events of the type (\ref{bbph})--(\ref{WWph}) contributions due
to `production radiation' from those due to `decay radiation', provided
significant differences exist between the corresponding differential
spectra. In particular, these can be exploited 
in order to enrich the experimental sample of events with photons
not emitted during the Higgs decays.
Also, their knowledge could eventually be useful in disentangling
Higgs signal from background processes (both irreducible and reducible ones).

Finally, the cut on the transverse momentum of the photon
has actually also been applied to all other particles in the final states
of reactions (\ref{bbph})--(\ref{WW}), apart from the neutrinos.
Such a constraint should generally meet the
requirements due to the finite coverage in energy and polar angle
of the NLC detectors.

\subsection*{3.1. The Higgs process 
$e^+e^-\ar\bar\nu_e\nu_e b\bar b\gamma$}

We first consider the $b\bar b$ Higgs decay channel.  The cross sections for
processes (\ref{bbph}) and (\ref{bb}) are given in Fig.~3, for the
mentioned choice of CM energies. They are presented as a function
of the Higgs mass $\MH$. 
For illustrative purposes we have considered the mass range
60 GeV $\Ord\MH\Ord 350$ GeV, however the largest rates occur for
$\MH\Ord140$ GeV (intermediate mass range), where the $H\ar b\bar b$ decay
has the largest branching ratio (BR). For $\MH\OOrd 140$ GeV the
off-shell channel $H\ar W^*W^*$ starts dominating the \sm\ Higgs decay
phenomenology. This is reflected in the steep decrease of the 
cross sections, both at leading and next-to-leading order, slightly before
the real $2\MW$ threshold.

The main feature of Fig.~3 is certainly the relatively large value
of the rates for the radiative process (\ref{bbph}) compared to those
of the non-radiative reaction (\ref{bb}). In fact, the former is at least
$10\%$ of the latter over all the interesting range of $\MH$, at all
values of $\sqrt s$. It increases with the CM energy and it is
largely independent of the Higgs mass.
In particular, if one assumes an integrated luminosity of $\int{\cal L}dt
=10~\mathrm{fb}^{-1}$, more than 300 events 
with hard photons  
could be produced per year, at a NLC with $\sqrt s=1000$ GeV  and for
$\MH\approx60$ GeV.
Such a number gets smaller with decreasing CM energy and 
increasing Higgs mass.

The kinematic properties of events of the type  (\ref{bbph}) and (\ref{bb})
are illustrated in Figs.~5a--c, for $\sqrt s=300, 500$ and 1000 GeV,
respectively. The selection of Higgs masses we have chosen here is
$\MH=60, 100$ and 140 GeV. For these values, the Higgs width is 
rather small, $\Gamma_H\Ord8$ MeV, such that hard photon emission in the
decay process $H\ar b\bar b\gamma$ (diagrams 3--4 in Fig.~1)
is heavily suppressed. In fact, the contribution due to the two mentioned
diagrams to the total cross section in 
$e^+e^-\rightarrow  \bar\nu_e\nu_e H(\gamma)\ar \bar\nu_e\nu_e b\bar b\gamma$
events is rather small (by more than one order of
magnitude, for all $\sqrt s$ and $\MH$ combinations)
compared to the contribution
due to graphs 1--2 \& 5--6 in Fig.~1, that is when the photon 
is emitted by the electron/positron and gauge boson lines. 
This is clearly reflected in the NLO $M_{b\bar b}$ invariant mass distribution
(upper left in Figs.~5a--c). In fact, the smearing towards low
masses of the Higgs
peaks due to the photon emission in the radiative $H\ar b\bar b\gamma$ decay
is essentially irrelevant for realistic phenomenological analyses: at the
level of ${\cal O}(10^{-2})$ or even less\footnote{Note the logarithmic
scale in all the figures.}. 

The suppression of the `decay radiation' is also visible in the
invariant mass of the $b\bar b\gamma$ system (upper right plots in
Figs.~5a--c). In the sense that the corresponding spectra show
a step (rather than 
a peak) at $M_{b\bar b\gamma}\approx\MH$, with a long and 
quantitatively relevant tail 
for $M_{b\bar b\gamma}>\MH$, due to the photons from the `production 
radiation'. In practise, the latter largely `overwhelm' those 
originating in the `decay radiation', which would produce a Breit-Wigner peak.

The distribution in energy of the radiated photon is displayed 
in the bottom left plots of the same figures. It is qualitatively
the same regardless of the actual value of the Higgs mass.
In particular, very hard photons are suppressed at smaller 
(see Fig.~5a) and enhanced at larger   (see Fig.~5c) CM energies, while
the dependence of the spectrum at 500 GeV (see Fig.~5b) is 
roughly exponential.
An additional suppression comes with the increase of the Higgs mass, at
fixed $\sqrt s$, especially if $\sqrt s\le 500$ GeV, whereas for $\sqrt s=1000$
GeV such an 
effect is quite small. Furthermore, as one expects, the `decay
radiation' photons are always softer than the `production radiation' ones.

The distribution in missing transverse momentum is rather similar
for both processes (\ref{bbph}) and (\ref{bb}) (bottom right of the 
mentioned figures) and for the two radiative components at NLO (for this
reason the latter are not shown separately here). In general,
the $p_T^{\mathrm{miss}}$ spectrum is
only slightly harder for the non-radiative process.
The maximum of the distribution is more pronounced as the CM energy
increases and does not depend on the values of $\MH$.

\subsection*{3.2. The Higgs process 
$e^+e^-\ar\bar\nu_e\nu_e \mathrm{jjjj}\gamma$}

Integrated cross sections at the CM energies $\sqrt s=300,500$ and 1000 GeV
for processes (\ref{WWph}) and (\ref{WW}) are shown in Fig.~4, as a 
function of the Higgs mass in the heavy mass range\footnote{We consider 
values of $\MH$ (greater than $180$ GeV) far above the real
$WW$ threshold at $2\MW\approx160$ GeV, in order to avoid complications
due to having adopted here a narrow width approximation for the
$W$ boson, when the real value of $\Gamma_W$ is instead around 2 GeV.}.
Whereas at lower values of $\sqrt s$ (upper plot) the production of a 
Higgs scalar with $\MH\OOrd180$ GeV is heavily suppressed,
 at larger CM energies (central and lower 
plots) rates are high enough to produce Higgs bosons up to  masses of 
450 GeV or so. 
In particular, at $\sqrt s=1000$ GeV,
cross sections for both processes (\ref{WWph}) and (\ref{WW}) diminish
by only a factor of 8 if $\MH$ increases from 180 to 450 GeV.
In contrast, at $\sqrt s=300$ GeV, there is a steep decrease of the rates
as $\MH$ approaches
$\sqrt s$. Curiously,
in this case, the shape of the curves is rather similar to those
of processes (\ref{bbph}) and (\ref{bb}) (Fig.~3 upper plot). 
However, there the effect was due to the Higgs branching ratio,
whereas here it occurs because of a phase space suppression 
on the production mechanism. 
Again, the most remarkable feature of Fig.~4 is a large rate for the 
radiative reaction (\ref{WWph}), compared to the lowest order one (\ref{WW}).
This is generally true for all relevant combinations
of masses and energies.
The ratio between the rates of the two processes
is less than a factor of 10, and 
up to 100 radiative events per year can be expected (lower plot for
the minimum Higgs mass). 

Figs.~6a--c show the kinematic properties of the
particles in the finale state of reactions (\ref{WWph}) and (\ref{WW}).
Since in the heavy mass range the Higgs width is quite large (it varies from
$\approx0.6$ GeV at 180 GeV to $\approx40$ GeV at 460 GeV !), hard
photon radiation
in the Higgs decay is no longer suppressed (unlike the
case of $H\ar b\bar b\gamma$
radiative decays), especially at large Higgs masses.
Moreover, in process (\ref{WWph}), there
are six graphs which contribute to produce a peak in the 
invariant mass of the $\mathrm{jjjj}\gamma$ system (i.e., number 3--4 \& 7--10
in Fig.~2), since
in the decay process $H\ar WW\ar(\gamma) \mathrm{jjjj}\gamma$ 
photon emission can
take place off two boson and four fermion lines. 
These two aspects (i.e., large Higgs width and `multiple' emission)
act in such a way that in the end the 
amplitude squared corresponding to the contribution of the six
mentioned diagrams 
(`decay radiation') is comparable to that of the
others graphs (`production radiation'), over all the $\MH$ spectrum
and for all values of $\sqrt s$ considered here. 
This can be clearly appreciated by looking at 
the $M_{\mathrm{jjjj}\gamma}$ spectra (Figs.~6a--c, upper right). 
All the distributions in invariant mass of the $\mathrm{jjjj}\gamma$ system
show a clear peak 
around the selected values of $M_H$ (upper right plots in 
Figs.~6a--c). When the Higgs boson is lighter, such that its 
width is relatively narrow (for 180 GeV in the plots), 
many hard photons still come from the `production radiation', giving the
long tail especially 
visible in the continuous lines, similarly to process (\ref{bbph}). However,
contrary to the case of $H\ar b\bar b\gamma$ resonant
decays, here the peaks at $M_{\mathrm{jjjj}\gamma}\approx\MH$ remain
visible.
When instead the Higgs width is larger (for $\MH\OOrd220$ GeV in the plots),
a large part of the photons come from the `decay radiation', 
such that they contribute
to build up a rather clear resonant dependence and the tail
at $M_{\mathrm{jjjj}\gamma}> M_H$ starts 
disappearing (dotted and dashed lines).
It is worth noticing that a phase space effect also 
contributes to enhance the above suppression
at large values of $M_{\mathrm{jjjj}\gamma}$, such that in the end, in some
instances (for example, at $\sqrt s=300$ GeV and $\MH\OOrd220$ GeV, Fig.~6a),
the NLO $M_{\mathrm{jjjj}\gamma}$ and $M_{\mathrm{jjjj}}$ 
distributions look pretty similar. 

When considering the case of the distributions 
in $M_{\mathrm{jjjj}}$ at NLO, effects due to phase space 
enhancement/suppression
play a more determinant r\^ole. In the sense that, when the portion of phase 
space available is restricted (that is at $\sqrt s=300 $ GeV, Fig.~6a
upper left, the shaded curves) the main
effect due to the hard photons is the appearance of a tail at 
$M_{\mathrm{jjjj}}<\MH$ (see especially the cases $\MH=220$ and 260 GeV).
This is a consequence of a superposition of a rather symmetrical resonance
(due to the `production radiation' diagrams, whose shape looks similar to the
one at LO, appropriately rescaled) and of a spectrum (due to the `decay
radiation' diagrams) that is strongly shifted towards low invariant 
masses\footnote{Note
that the appearance of a low mass `tail' also for the LO distribution 
in $M_{\mathrm{jjjj}}$ when $\MH=260$ GeV
at $\sqrt s=300$ GeV (upper dashed line in Fig.~6a, upper left plot) 
is accidental, as
in this case the spectrum is close to its upper kinematic boundaries.}.
As $\sqrt s$ increases the low mass tail effect due 
to diagrams 3--4 \& 7--10
in Fig.~2 is counterbalanced by the fact that the contributions
at NLO to the $M_{\mathrm{jjjj}}$ spectra due to diagrams 1--2 and 5--6 
in Fig~2 have no longer a symmetrical shape (compare to 
the same distributions at LO, especially in Fig.~6c),
but this is significantly shifted towards high masses. 
In many cases, in fact, the
phase space available for $M_{\mathrm{jjjj}}>\MH$ is much larger than that
in the complementary region $M_{\mathrm{jjjj}}<\MH$.
Therefore, for $\sqrt s\OOrd 500$ GeV (Figs.~6b--c), the overall effect is
a rather symmetrical resonance at NLO too. 
The same effect was not quantitatively appreciable in the case of process
(\ref{bbph}), since there 
the small width of the Higgs boson in the intermediate mass range drastically
reduced the relevance of configurations
in which $M_{b\bar b}-M_{H}\gg \Gamma_H$.
 
The energy spectra of the emitted photon are shown in the bottom left
plots of Fig.~6a--c. Again, one can see that very hard photons
are not allowed if the Higgs mass/width
and the CM energy are small (continuous line in 
Fig.~6a) and that photons created by the `production radiation' are in
most cases
harder than those generated by the `decay radiation', although the difference
is less evident than for process (\ref{bbph}). Unfortunately, this is not
true at $\sqrt s=300$ GeV, where the distribution of the full sample
closely resembles in shape that produced by diagrams 3--4 \& 7--10 in Fig.~2.

The behaviour of the spectra in $p_T^{miss}$ (bottom right frames
of Fig.~6a--c) shows that, also in the case  of processes (\ref{WWph}) and 
(\ref{WW}), the average value of the missing transverse momentum is
approximately the same both at leading and at next-to-leading order.
Moreover, if one studies separately the contributions to the NLO
spectra due to the `production radiation' and `decay radiation'
diagrams, one realises that once again they do not show any 
substantial difference, their shape being described by the lower curves
in Fig.~6a--c, with appropriate normalisations.

Finally, a last comment is in order if one considers that a heavy $H$ boson
can decay into four jets also via the $ZZ$ channel. Since 
$\mathrm{BR}(W\ar\mathrm{jj})\approx\mathrm{BR}(Z\ar\mathrm{jj})$
(when also $b$-jets are considered in the $Z$ decay), 
the contribution to the total rates of the $2\ar6$ leading order reaction
$e^+e^-\rightarrow\bar\nu_e\nu_e H \ar \bar\nu_e\nu_e [WW+ZZ]  
\ar \bar\nu_e\nu_e \mathrm{jjjj}$ due to diagrams involving the
$H\ar ZZ$ decay is proportional to the corresponding 
branching ratio. In particular, in the heavy mass range
investigated here, these are expected to be smaller by a factor of
$\approx 20$, for $\MH=180$ GeV, and of $\approx2.5$, for $M_H\OOrd250$ GeV,
compared to rates due to contributions proceeding through
the $H\ar WW$ channel.
At next-lo-leading order, that is for the $2\ar7$ reaction 
$e^+e^-\rightarrow\bar\nu_e\nu_e H (\gamma) \ar \bar\nu_e\nu_e [WW+ZZ](\gamma) 
\ar \bar\nu_e\nu_e \mathrm{jjjj}\gamma$, a further
suppression occurs. In fact, when $H\ar ZZ\ar\mathrm{jjjj}\gamma$, 
diagrams 3--4 in Fig.~2 no longer appear in the matrix element.
However, the contributions of the corresponding squared amplitudes
to the total cross sections of the above process 
are smaller than those of graphs 1--2
\& 5--6, as well as than those of graphs 7--10
(Fig.~2). It varies from a few percent at small
energies (i.e., $\sqrt s=300$ and 500 GeV) and small Higgs masses,
up to $\approx15\%$ for $\sqrt s=1000$ GeV and large values of 
$\MH$\footnote{We also expect an additional suppression for diagrams 7--10
of Fig.~2 when $\MH\Ord 2M_Z$.}. Thus,
in the end, the inclusion into the calculations of diagrams 
involving $H\ar ZZ\ar\mathrm{jjjj}(\gamma)$ decays could in some
instances reduces the relative importance of events with hard photons.
Conversely, this also indicates that experimental procedures exploiting
the $H\ar\mathrm{jjjj}$ decay inclusively, or the $H\ar ZZ\ar \mathrm{jjjj}$
on its own, are more 
likely to be less sensitive to higher order electromagnetic
effects.

\section*{4. Summary and conclusions}

In this paper we have studied the two radiative Higgs processes
$$
e^+e^-\rightarrow\bar\nu_e\nu_e H (\gamma)\ar \bar\nu_e\nu_e b\bar b \gamma,
$$
$$
e^+e^-\rightarrow\bar\nu_e\nu_e H (\gamma)\ar \bar\nu_e\nu_e WW (\gamma)
\ar \bar\nu_e\nu_e \mathrm{jjjj}\gamma.
$$
We have considered the case of hard and detectable photons, by adopting
the cuts $p_T^{b,~\mathrm{j},\gamma}>1$ GeV, on the transverse momentum
of all particles in the final states, and 
$\cos\theta_{b\gamma,~\mathrm{j}\gamma}<0.95$, for the separation
between the photon and the hadronic systems.
We have then compared integrated and differential rates obtained
for the two above processes to those of the reactions 
($p_T^{b,~\mathrm{j}}>1$ GeV)
$$
e^+e^-\rightarrow\bar\nu_e\nu_e H \ar \bar\nu_e\nu_e b\bar b,
$$
$$
e^+e^-\rightarrow\bar\nu_e\nu_e H \ar \bar\nu_e\nu_e WW 
\ar \bar\nu_e\nu_e \mathrm{jjjj}.
$$
These four processes represent the main tree-level 
production and decay mechanisms
of a Standard Model Higgs boson in the $WW$-fusion channel, at the
centre-of-mass energies typical of an $e^+e^-$ Next Linear Collider,
at next-to-leading and leading order in the electromagnetic coupling
constant, respectively. The final states $ \bar\nu_e\nu_e b\bar b(\gamma)$
have been considered in order to give account of the most likely
decay mechanism of the Higgs boson in the intermediate mass range
$\MH\Ord2\MW$ (into a $b\bar b$ pair), whereas the  
$\bar\nu_e\nu_e \mathrm{jjjj}(\gamma)$ ones refer to the favourite Higgs
signature in the heavy mass range $M\OOrd2M_W$ (into four jets via $WW$). 
The values 300, 500 and 1000 GeV have been considered for $\sqrt s$.

Our study has been motivated by noticing that
 the presence of hard electromagnetic
emission via bremsstrahlung photons from the initial state is somewhat
unavoidable at these machines. In fact, whereas at
current lower energy leptonic colliders (such as LEP1 and
SLC, and partially LEP2 as well) the width of the unstable particles 
produced in the $e^+e^-$ direct annihilation subprocess ($Z$ and $WW$, 
respectively) naturally imposes a cut-off on hard photon radiation from
the incoming electron and positron lines, 
this is no longer the case at the Next Linear Collider.
Therefore one inevitably has to deal with such electromagnetic effects
when attempting phenomenological analyses. Furthermore, in experimental
data samples, radiative events in which the photon comes from the Higgs 
production mechanism
are not separable from those in which it comes from the Higgs decay
channels. Whereas in the former the shape of the Higgs resonances
is generally not spoiled by hard photons, in the second this can in 
principle be heavily distorted. 

Since in our computation the contributions due to the loop diagrams
were missing, we have not performed a summation through the 
orders ${\cal O}(\alpha_{\mathrm{em}}^5)$ and 
${\cal O}(\alpha_{\mathrm{em}}^7)$.
Certainly, definite estimates of the ${\cal O}(\alpha_{\mathrm{em}})$ effects
should be given only after a complete 
calculation including virtual photons contributions.
However, the following interesting indications can be extracted from our 
partial results. 
\begin{itemize}
\item Rates at next-to-leading order involving hard photons can be up to 
$30\%$ of those at leading order. In particular, for all the relevant Higgs
masses and CM energies considered here, these are never smaller than $10\%$.
This corresponds to a rather large effect, in some cases
comparable to or even larger than those due to
beam related phenomena (such as beamsstrahlung, Linac
energy spread), which are expected to take
place at the high energy $e^+e^-$ colliders of the next generation.
\item The Higgs width acts as a regulator of the size of the EM emission
in the Higgs decay processes. The smaller it is, the more suppressed 
the contribution to the total rates of diagrams involving photon
radiation after Higgs production. Therefore, the smearing
of the resonant distributions can be quantitatively significant
only in case of Higgs masses in the heavy range (i.e., $\MH\OOrd 2M_W$).
This should remain true also for poor detector resolutions.
\item If one considers heavy Higgs bosons, then two different scenarios appear.
\end{itemize}

\begin{enumerate}
\item For a NLC with $\sqrt s\OOrd500$ GeV, the higher order
distributions relevant to
Higgs searches and studies
(i.e., the invariant masses of the Higgs decays products $b\bar b$ and
$\mathrm{jjjj}$) generally 
have a shape similar to that of the $2\ar4$ and $2\ar6$ leading
processes. This is because
the smearing of the Higgs peaks towards low masses
due to the emission of hard photons in the Higgs decays is counterbalanced
by a tail towards large masses due to the large portion of phase space
available to photons emitted in the Higgs 
production mechanism. In this case, the inclusion of the complete 
${\cal O}(\alpha_{\mathrm{em}})$ corrections is merely a matter of a different
normalisation of the lowest order distributions.
\item In contrast, for a $\sqrt s=300$ GeV NLC, where such phase
space effects are negligible, the resonant differential spectra
at NLO do show a quantitatively significant
low mass tail. The effect clearly increases
with the Higgs mass/width. 
In this case, as rates of events with hard photons can be large, a complete
${\cal O}(\alpha_{\mathrm{em}})$ calculation is desirable in order to
assess the correct shape of the distributions at the complete NLO.
\end{enumerate}
\begin{itemize}
\item Since events with hard and detectable photons are defined on their
own once appropriate cuts are applied to remove the infrared 
divergences (and these can well
coincide with those dictated by the tagging procedures of the experimental
analyses), it is feasible to study such events separately. An interesting
features of these is that the invariant masses of the systems $b\bar b\gamma$
(for intermediate mass Higgses) 
and especially $\mathrm{jjjj}\gamma$ (for heavy mass Higgses)
show a clear step/peak around the actual value of $\MH$.
In many cases, a detectable number of events in the resonant region can be 
produced 
after a few years of running at the nominal collider luminosity.
\item Distributions 
which could naturally allow a separation between radiative events 
with photons emitted in the production (on the one hand) 
and  decay (on the other hand) stages (such as,
for example, the energy of the photon and the transverse missing
momentum of the neutrino pair) are not generally helpful. The 
$p_T^{\mathrm{miss}}$ spectrum 
presents the same behaviour for all energies and 
Higgs masses, both for the leading process and the two
next-to-leading radiative contributions. 
The $E_\gamma$ spectrum 
could be exploited in most cases, but unfortunately not
when $\sqrt s=300$ GeV and $\MH\OOrd2\MW$, since in this case
the spectra of the `production radiation' and of the `decay radiation'
are extremely similar.
\end{itemize}

\noindent
In general, we stress that,
in order to counterbalance the positive
infrared divergences of soft and collinear (to both electron/positron 
and partons in the initial and final states, respectively) real EM emission,
the corrections due to virtual photons in loop diagrams 
should generally be negative and could well reduce the relative contributions
of events with the kinematics of the lowest-order processes.
Therefore, 
the hard photon effects predicted here could well be enlarged by the
complete result.

In this paper, we have also pointed out that in four-jet decays 
of heavy Higgs bosons proceeding through the $ZZ$ channel (i.e., $H\ar
ZZ\ar \mathrm{jjjj}\gamma$) hard photon effects are smaller than
in the $H\ar WW(\gamma)\ar \mathrm{jjjj}\gamma$ case, but only over 
appropriate regions of the $\MH$ range (that is,
well above the $2\MZ$ Higgs decay threshold).
 
Finally, some technical details.
Our computations have been performed by resorting to a numerical evaluation
of the exact matrix elements of all the above processes, by means of
helicity amplitude techniques, integrating them 
over the appropriate phase spaces, using a standard Monte Carlo routine.
A dedicated treatment of the various resonant diagrams makes the
{\tt FORTRAN} code produced rather accurate and fast.
In this respect, we emphasise that we have used no approximations here,
apart from that of constraining the two $W$ bosons entering in the Higgs
decay channel $H\ar WW(\gamma)\ar \mathrm{jjjj}(\gamma)$ to be on-shell.
This has been done in order to reduce the total amount of CPU usage, especially
needed when performing a $N=3n-5$ (assuming no transverse polarization
of the beams) dimensional integration for an $n$-particle final state. However,
the knowledge of the off-shell decay phenomenology of the Higgs boson into
two $W$'s makes straightforward the extrapolation of our results
to the relevant $\MH$ range\footnote{Alternatively, the constraints on the 
invariant masses flowing through the $W$-resonances can be removed, and 
the integration performed over the original $N$ dimensions, at the cost
of additional CPU time.}.
Furthermore, some of the interferences between Feynman diagrams with
different resonant structures have been neglected in the presentation
of our results, as these are numerically negligible for the values of
$\sqrt s$ and $\MH$ adopted here.

\section*{Acknowledgments}

We are grateful to Gavin Salam for reading the manuscript.
We thank the UK PPARC for support. This work is financed in part by the
Ministero dell' Universit\`a e della Ricerca Scientifica and  by the EC
Programme ``Human Capital and Mobility'', contract CHRX--CT--93--0357
(DG 12 COMA).






\subsection*{Figure Captions}

\begin{description}

\item[Fig.~1 ] Feynman diagrams contributing at tree-level to process 
(\ref{bbph}).

\item[Fig.~2 ] Feynman diagrams contributing at tree-level to process 
(\ref{WWph}).

\item[Fig.~3 ] Cross sections of the processes 
(\ref{bbph}) and (\ref{bb}) as a function
of the Higgs mass, at 
$\sqrt s =300$ GeV (upper plot),
$\sqrt s =500$ GeV (central plot),
$\sqrt s =1000$ GeV (lower plot), 
after the cuts $p_T^{b,\gamma}>1$ GeV and $\cos\theta_{b\gamma}<0.95$.

\item[Fig.~4 ] Cross sections of the processes 
(\ref{WWph}) and (\ref{WW}) as a function
of the Higgs mass, at 
$\sqrt s =300$ GeV (upper plot),
$\sqrt s =500$ GeV (central plot),
$\sqrt s =1000$ GeV (lower plot), 
after the cuts $p_T^{~\mathrm{j},\gamma}>1$ GeV and 
$\cos\theta_{\mathrm{j}\gamma}<0.95$.

\item[Fig.~5 ] Distributions in invariant mass of the $b\bar b$ pair
(upper left plot), in invariant mass of the $b\bar b\gamma$ system
(upper right plot), in energy of the photon (lower left plot)
and in missing transverse momentum (lower right plot) for
processes (\ref{bbph}) and (\ref{bb}), after the cuts 
$p_T^{b,\gamma}>1$ GeV and $\cos\theta_{b\gamma}<0.95$, for a selection
of Higgs masses:
({\bf a}) at $\sqrt s=300$ GeV (bins of 2 GeV); 
({\bf b}) at $\sqrt s=500$ GeV (bins of 4 GeV); 
({\bf c}) at $\sqrt s=1000$ GeV (bins of 5 GeV).
In the first and third plot (clockwise) the
upper histograms refer to rates from the non-radiative process
(\ref{bb}), whereas the lower histograms correspond to rates from
the radiative process (\ref{bbph}). In the first plot the lowest
order rates are
shaded. In the fourth plot the
upper histograms refer to rates from the complete set of
diagrams describing process
(\ref{bbph}), whereas the lower histograms correspond to rates from
contributions due to `decay radiation' graphs only.

\item[Fig.~6 ] Distributions in invariant mass of the $4\mbox{jet}$ system
(upper left plot), in invariant mass of the $4\mbox{jet}\gamma$ system
(upper right plot), in energy of the photon (lower left plot)
and in missing transverse momentum (lower right plot) for
processes (\ref{WWph}) and (\ref{WW}), after the cuts 
$p_T^{~\mathrm{j},\gamma}>1$ GeV and $\cos\theta_{\mathrm{j}\gamma}<0.95$, 
for various selections of Higgs masses:
({\bf a}) at $\sqrt s=300$ GeV (bins of 2 GeV); 
({\bf b}) at $\sqrt s=500$ GeV (bins of 4 GeV); 
({\bf c}) at $\sqrt s=1000$ GeV (bins of 5 GeV).
In the first and third plot (clockwise) the
upper histograms refer to rates from the non-radiative process
(\ref{WW}), whereas the lower histograms correspond to rates from
the radiative process (\ref{WWph}). In the first plot the lowest
order rates are
shaded. In the fourth plot the
upper histograms refer to rates from the complete set of
diagrams describing process
(\ref{WWph}), whereas the lower histograms correspond to rates from
contributions due to `decay radiation' graphs only.

\end{description}
\vfill
\newpage

\begin{figure}[p]
~\epsfig{file=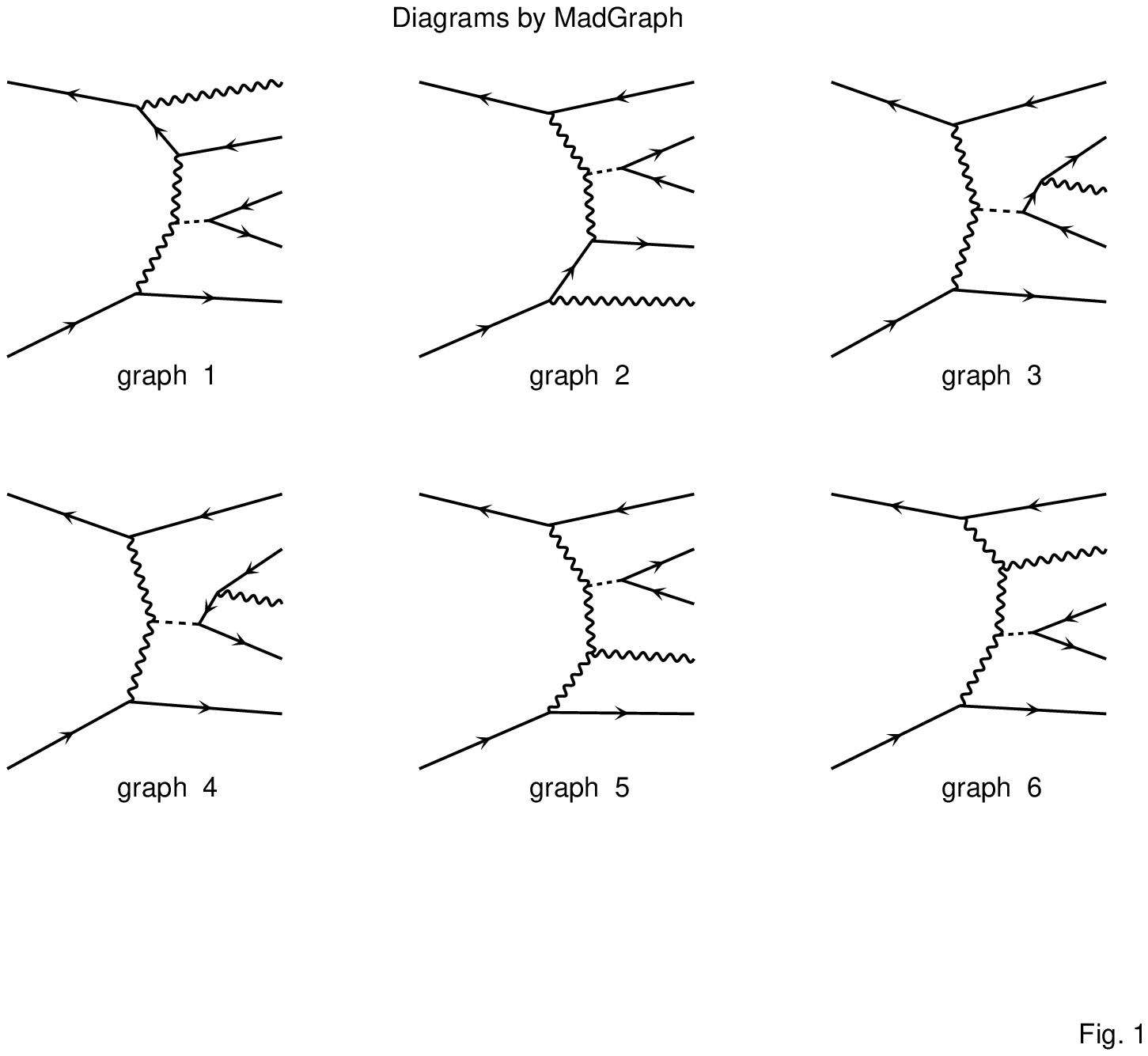,height=22cm}
\end{figure}
\stepcounter{figure}
\vfill
\clearpage

\begin{figure}[p]
~\epsfig{file=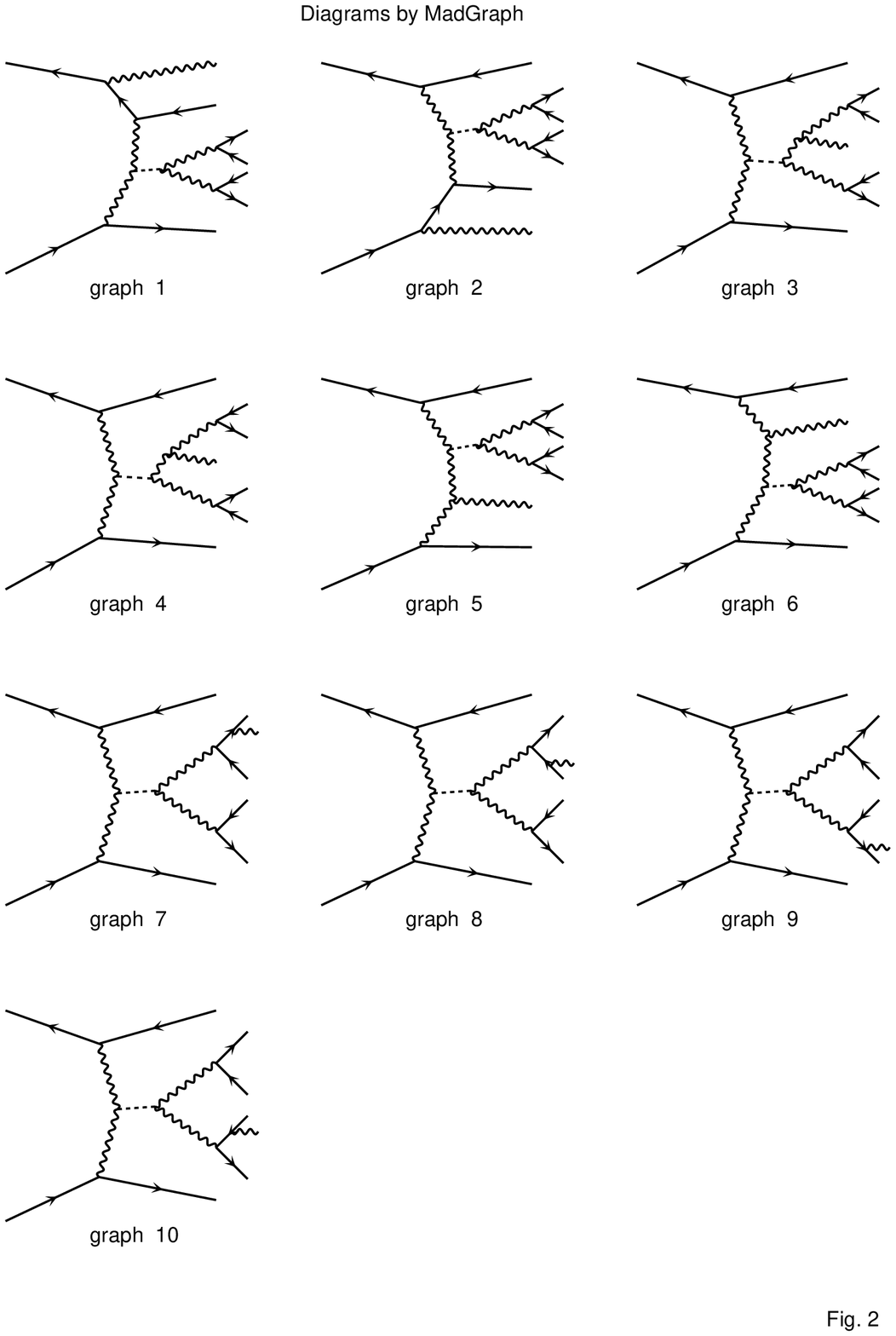,height=22cm}
\end{figure}
\stepcounter{figure}
\vfill
\clearpage

\begin{figure}[p]
~\epsfig{file=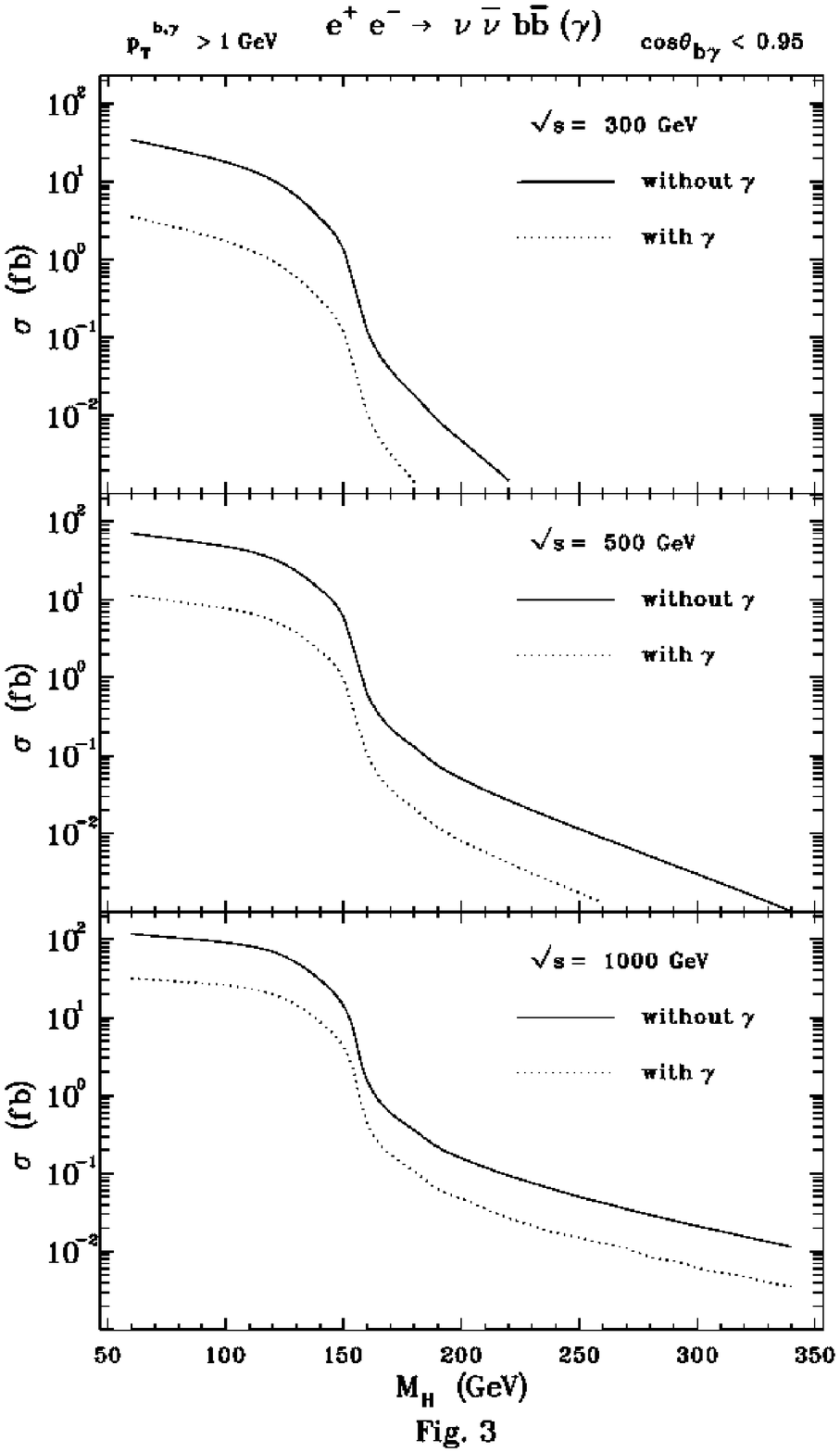,height=22cm}
\end{figure}
\stepcounter{figure}
\vfill
\clearpage

\begin{figure}[p]
~\epsfig{file=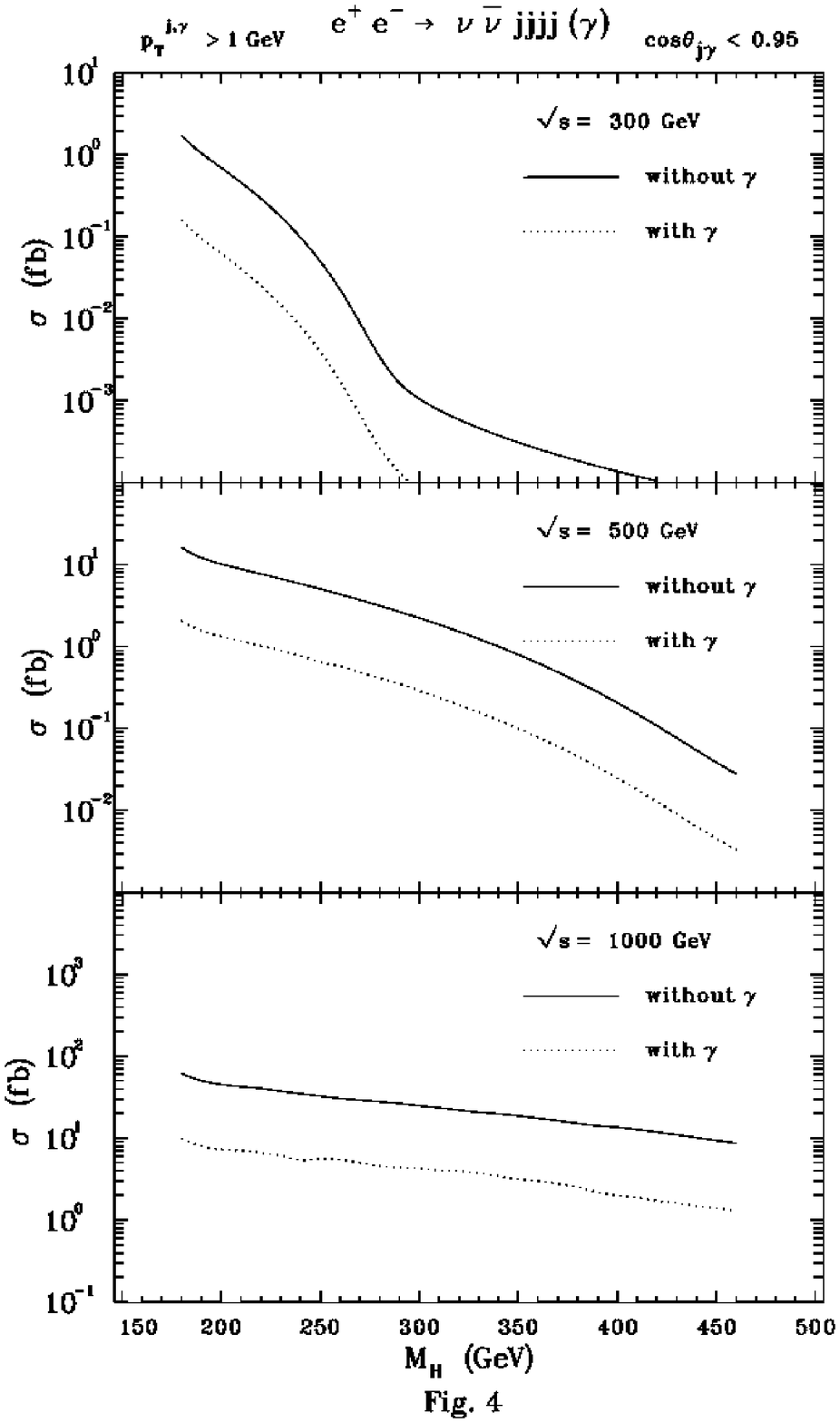,height=22cm}
\end{figure}
\stepcounter{figure}
\vfill
\clearpage

\begin{figure}[p]
~\epsfig{file=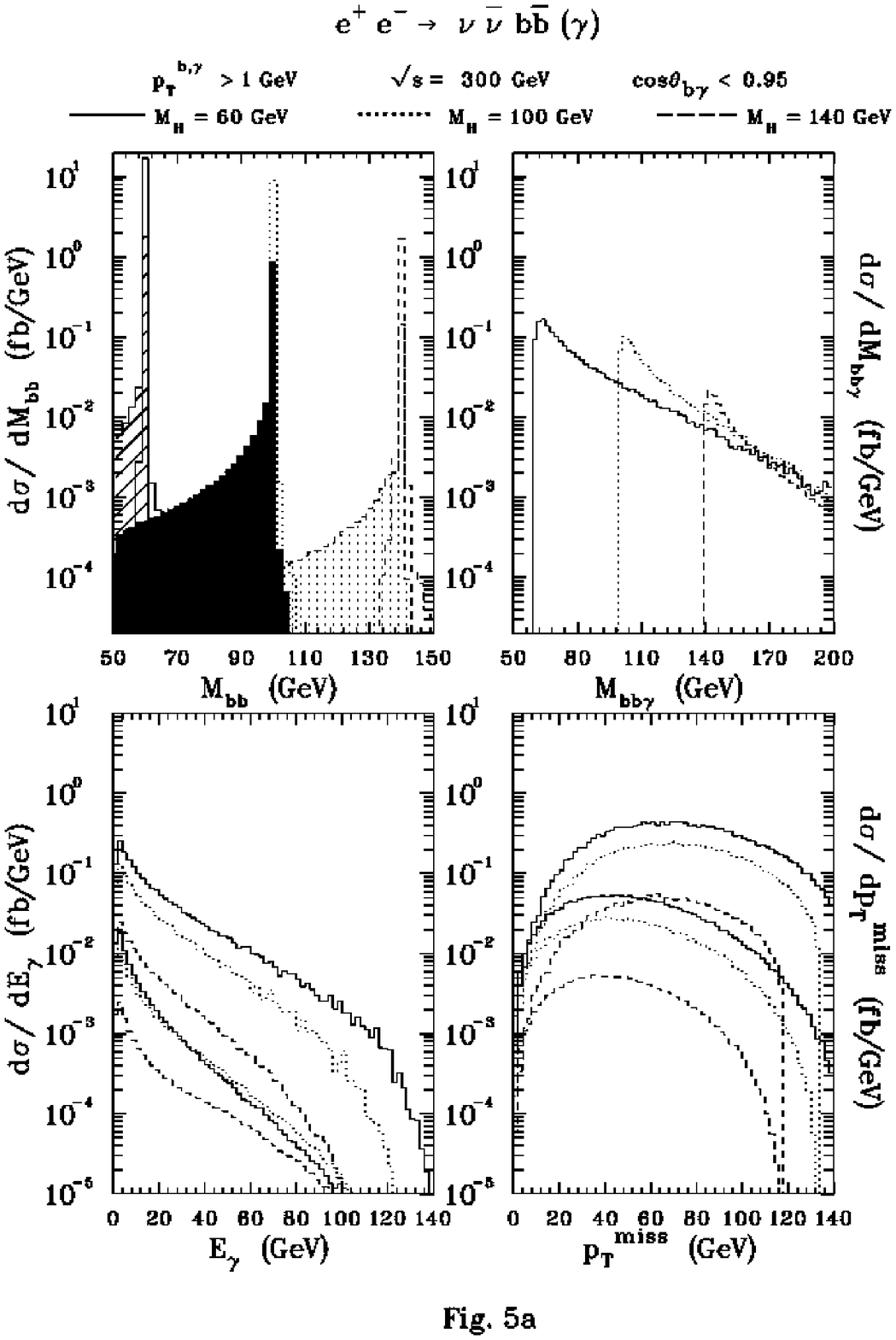,height=22cm}
\end{figure}
\stepcounter{figure}
\vfill
\clearpage

\begin{figure}[p]
~\epsfig{file=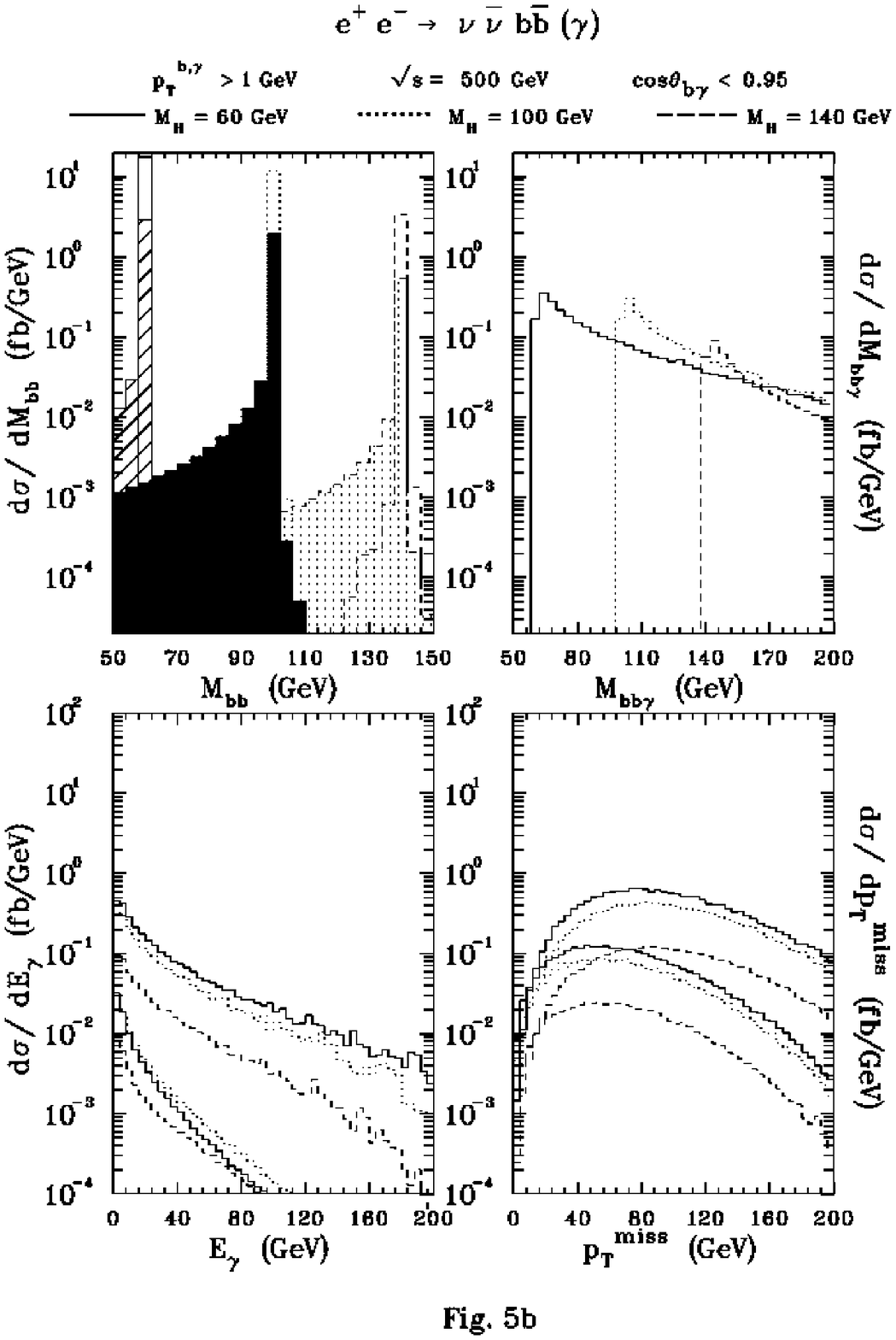,height=22cm}
\end{figure}
\stepcounter{figure}
\vfill
\clearpage

\begin{figure}[p]
~\epsfig{file=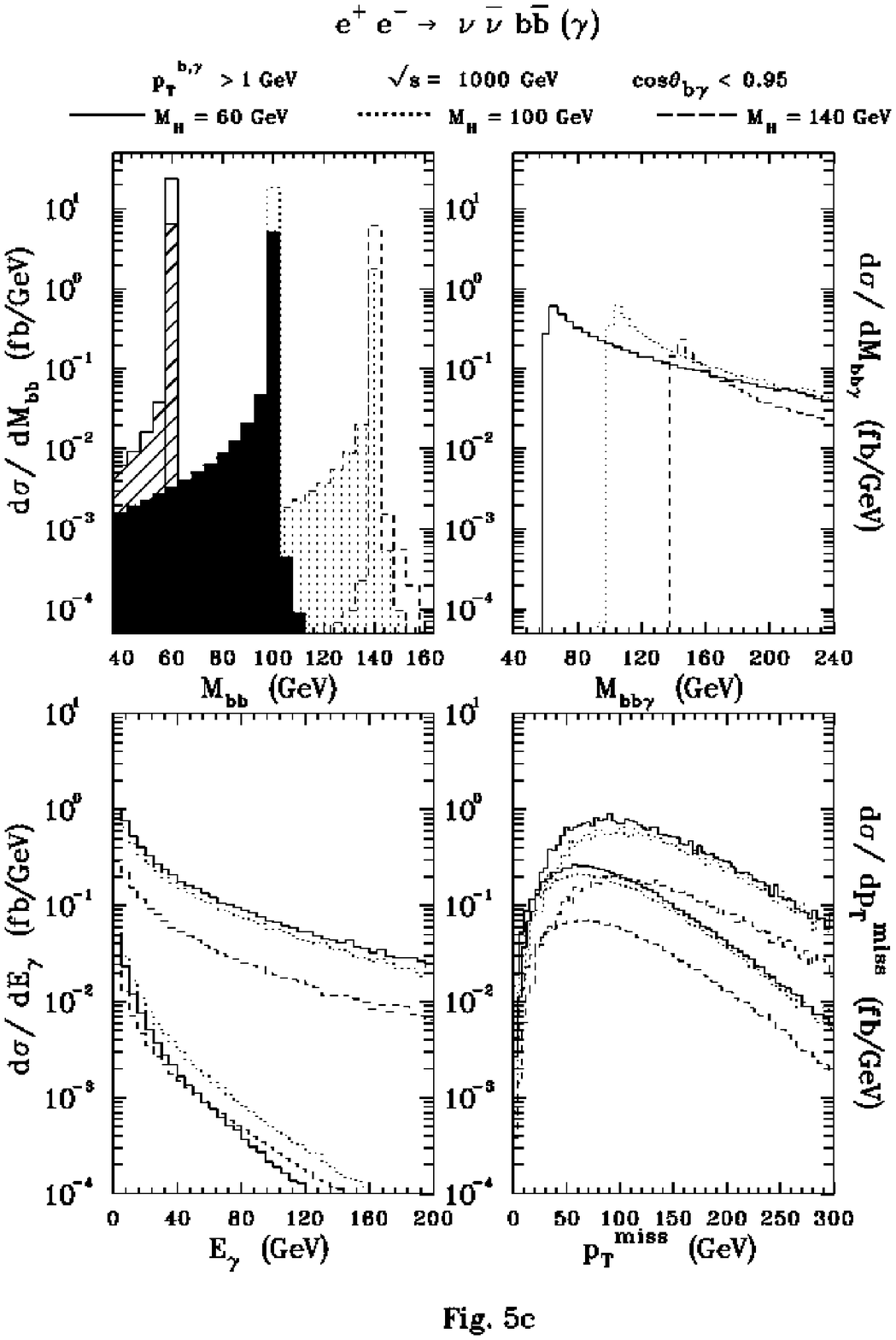,height=22cm}
\end{figure}
\stepcounter{figure}
\vfill
\clearpage

\begin{figure}[p]
~\epsfig{file=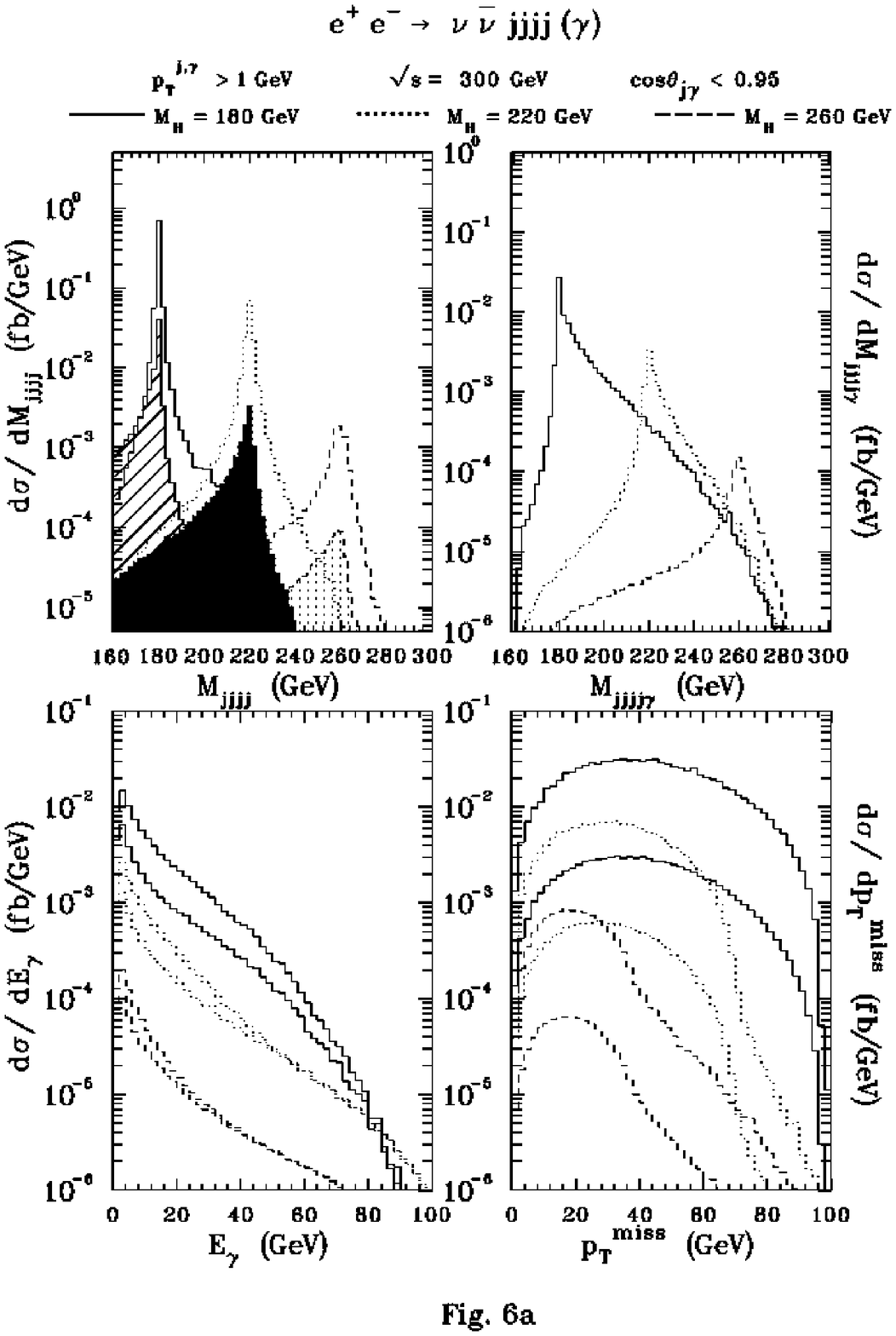,height=22cm}
\end{figure}
\stepcounter{figure}
\vfill
\clearpage

\begin{figure}[p]
~\epsfig{file=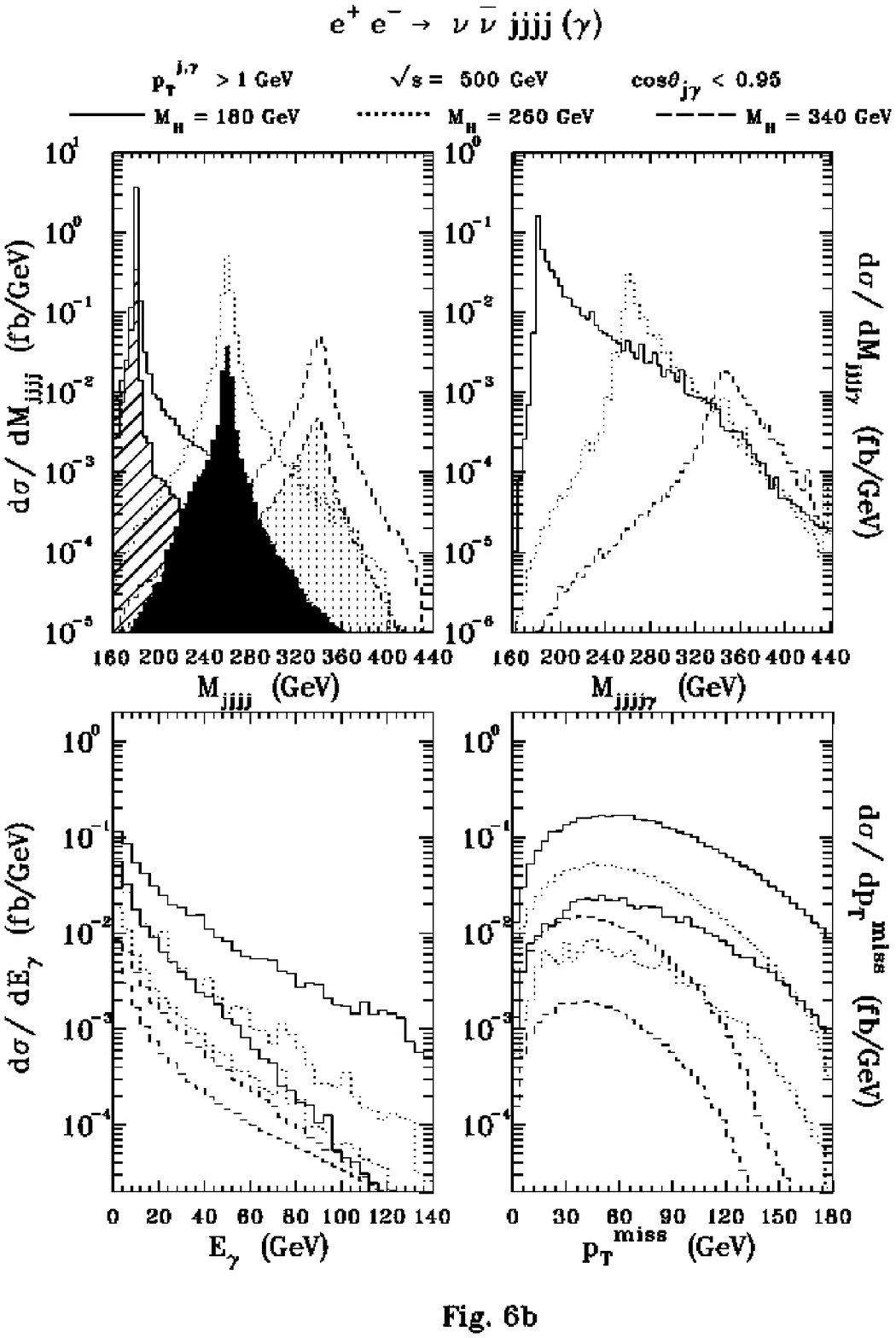,height=22cm}
\end{figure}
\stepcounter{figure}
\vfill
\clearpage

\begin{figure}[p]
~\epsfig{file=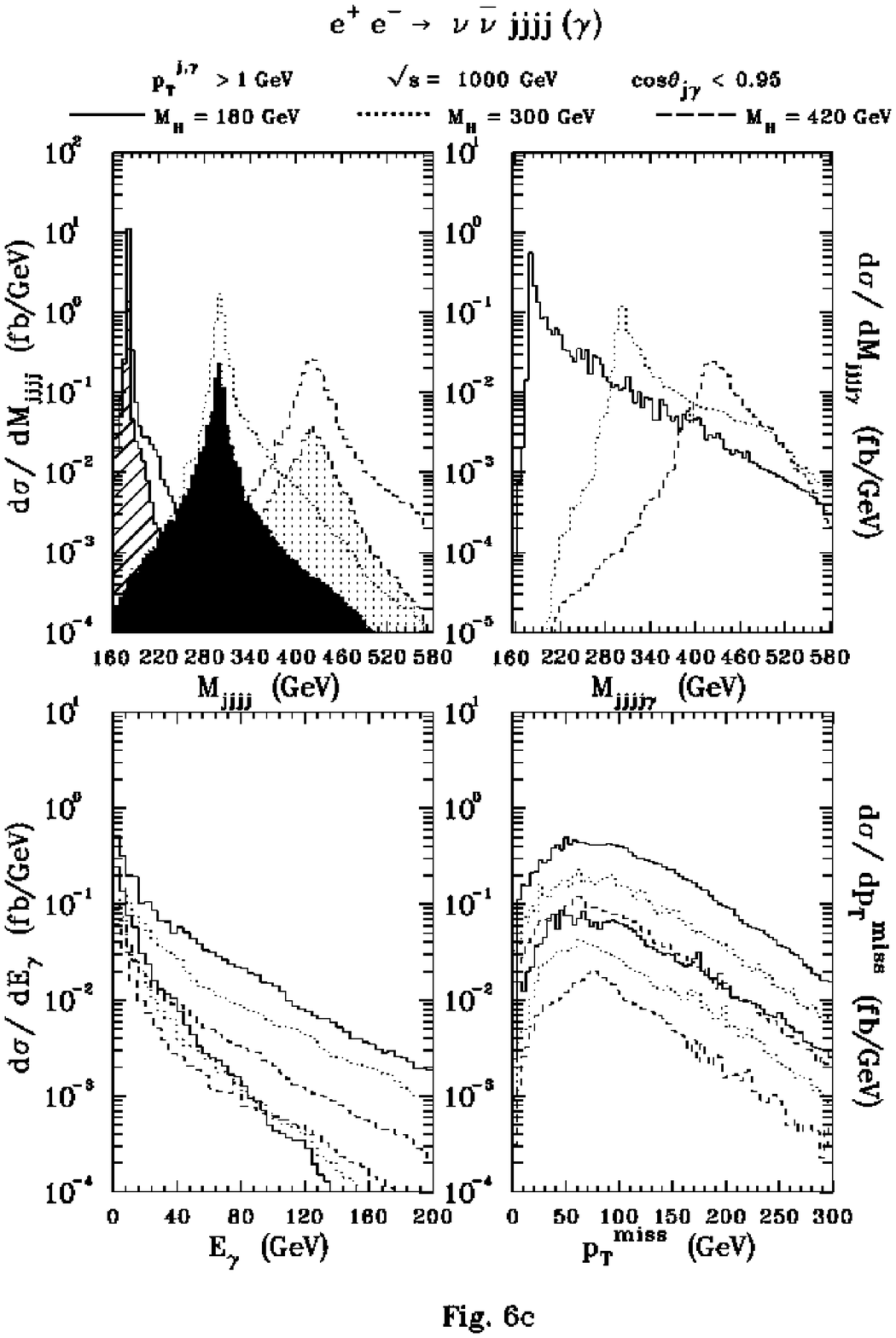,height=22cm}
\end{figure}
\stepcounter{figure}
\vfill
\clearpage

\vfill

\begin{thebibliography}{99}

\bibitem{epem1} Proc. of the ECFA workshop on LEP 200,
                A. Bohm and W. Hoogland eds.,
                Aachen FRG, 29 Sept.-1 Oct. 1986, CERN 87-08.

\bibitem{epem2} Proceedings of the Workshop
              ``{\it Phy\-sics and Ex\-pe\-ri\-men\-ts with Li\-ne\-ar
              Col\-li\-ders}'',
              Sa\-ar\-isel\-k\"a, Fin\-land, 9-14 Sep\-tem\-ber 1991,
              eds. R.~Orawa, P.~Eerola and M.~Nordberg,
              World Scientific Publishing, Singapore, 1992.

\bibitem{epem3} Proc. of the Workshop ``{\it $e^+e^-$ Collisions at
                500 GeV. The Physics Potential}\ '',
                Munich, Annecy, Hamburg, 3-4 February 1991, ed. P.M.~Zerwas,
                DESY pub. 92-123A/B,
                August 1992; DESY pub. 93-123C, December 1993.

\bibitem{epem4} Proc. of the ECFA workshop on ``{\it $e^+e^-$ Linear
               Colliders LC92}'', R.~Settles ed., Garmisch
               Partenkirchen, 25 July-2 Aug.
               1992, MPI-PhE/93-14, ECFA 93-154.

\bibitem{epem5} Proc. of the I-IV Workshops on Japan Linear Collider (JLC),
              KEK  1989, 1990, 1992, 1994,
              KEK-Reports 90-2, 91-10, 92-1, 94-1.

\bibitem{LHC} Proceedings of the ``{\it
Large Hadron Collider Workshop}'', Aachen, 4--9 October
1990, eds. G.~Jarlskog
and D.~Rein, Report CERN 90-10, ECFA 90-133, Geneva, 1990;\\
ATLAS Technical Proposal,
CERN/LHC/94-43 LHCC/P2 (December 1994);\\
CMS Technical Proposal, CERN/LHC/94-43 LHCC/P1 (December 1994).

\bibitem{bremSM} J.D.~Bjorken, Proceedings of the
                 ``{\it Summer Institute on Particle
                 Physics}'', {\it SLAC Report} 198 (1976);\\
                 B.W.~Lee, C.~Quigg and H.B.~Thacker, \pr D16 1977 1519;\\
                 J.~Ellis, M.K.~Gaillard and D.V.~Nanopoulos,
                 \np B106 1976 292;\\
                 B.L.~Ioffe and V.A.~Khoze,
                 {\it Sov. J. Part. Nucl.} {\bf 9} (1978) 50.

\bibitem{fusionSM} D.R.T.~Jones and S.T.~Petkov, \pl B84 1979 440;\\
                   R.N.~Cahn and S.~Dawson, \pl B136 1984 196;\\
                   K.~Hikasa, \pl B164 1985 341;\\
                   G.~Altarelli, B.~Mele and F.~Pitolli, \np B287 1987 205;\\
                   B.~Kniehl, {\it preprint} DESY 91-128, 1991.

\bibitem{DHKMZ} See for example:\\
A. Djouadi, D. Haidt, B.A. Kniehl, B. Mele and P.M. Zerwas, in 
Ref.~\cite{epem3}, part A, and references therein.

\bibitem{GHS} P.~Grosse-Wiesmann, D.~Haidt and H.J.~Schreiber, in
              Ref.~\cite{epem3}, part A.

\bibitem{ISR} T.~Barklow, P.~Chen and W.~Kozanecki, in 
Ref.~\cite{epem3}, part B, and references therein.

\bibitem{structure} F.A. Berends, W.L. van Neerven and G.J. Burgers,
{\it Nucl. Phys.} {\bf B297} (1988) 429; Erratum, {\it ibidem}
{\bf B304} (1988) 95;\\
E.A. Kuraev and V.S. Fadin, {\it Sov. J. Nucl. Phys.} {\bf 41}
(1985) 466;\\
G. Altarelli and G. Martinelli, Proceedings of the Workshop
`{\it Physics at LEP}', eds. J. Ellis
and R. Peccei, Geneva, 1986, CERN 86-02;\\
R. Kleiss, \np B347 1990 29;\\
O.~Nicrosini and L.~Trentadue, {\it Phys. Lett.} {\bf B196}
(1987) 551; {\it Z. Phys.} {\bf C39}  (1988) 479.

\bibitem{BCDKPZ} V.~Barger, K.~Cheung, A.~Djouadi, B.A.~Kniehl,
                   R.J.N.~Phillips and
                   P.M.~Zerwas, in  Ref.~\cite{epem3}, part C.

\bibitem{HZ} K.~Hagiwara and D.~Zeppenfeld,
{\it Nucl. Phys.} {\bf B274} (1986) 1.

\bibitem{HELAS} H.~Murayama, I.~Watanabe and K.~Hagiwara, HELAS: HELicity
                Amplitude Subroutines for Feynman Diagram Evaluations,
                {\it KEK Report} 91-11, January 1992.

\bibitem{tim} T.~Stelzer and W.F.~Long, {\it Comp. Phys. Comm.} {\bf 81}
              (1994) 357.

\bibitem{VEGAS} G.P.~Lepage, {\it Jour. Comp. Phys.} {\bf 27} (1978) 192.

\bibitem{cave9517} S. Moretti, \preprint\ DFTT 78/95, Cavendish--HEP--95/17,
revised February 1996 (to appear in {\it J. Phys.} {\bf G}), 
and references therein.

\bibitem{running} E.~Braaten and J.P.~Leveille, \pr D22 1980 715;\\
                  N.~Sakai, \pr D22 1980 2220;\\
                  T.~Inami and T.~Kubota, \np B179 1981 171;\\
                  M.~Drees and K.~Hikasa, \pl B240 1990 455;\\
                  S.G.~Gorishny, A.L.~Kataev, S.A.~Larin and
                  L.R.~Surguladze, \mpl A5 1990 2703;\\
                  L.R.~Surguladze, \pl B341 1994 60.

\bibitem{CST} D. Bardin, M. Bilenky, A. Olchevski and T. Riemann,
{\it Phys. Lett} {\bf B308} (1993) 403; Erratum, {\it ibidem} {\bf B357}
(1995) 725.

\bibitem{ISRcomplete} D. Bardin, D. Lehner and T. Riemann, \preprint\
DESY 96-028, February 1996.

\bibitem{KLN} T. Kinoshita, {\it J. Math. Phys.} {\bf 3} (1962) 650;\\
            T.D. Lee and M. Nauenberg, {\it Phys. Rev.} {\bf 133B} (1964) 1549.

\bibitem{Sudakov} V.V. Sudakov, {\it Sov. Phys. JETP} {\bf 3} (1956) 65. 

\end{thebibliography}
\end{document}